\documentclass[onecolumn,nofootinbib,showkeys]{revtex4}

\usepackage{graphicx}
\usepackage{amsmath}

\newcommand{\KMS}{\mbox{km s}^{-1}\,}

\def\ltsima{$\; \buildrel < \over \sim \;$}
\def\ltsim{\lower.5ex\hbox{\ltsima}}
\def\gtsima{$\; \buildrel > \over \sim \;$}
\def\gtsim{\lower.5ex\hbox{\gtsima}}

\def\KKF#1{Kidder kick formula#1}

\def\newacronym#1#2#3{\gdef#1{#3 (#2)\gdef#1{#2}}}

\newacronym{\NSF}{NSF}{National Science Foundation}
\newacronym{\NASA}{NASA}{National Aeronautics and Space Administration}
\newacronym{\lisa}{LISA}{the Laser Interferometer Space Antenna}
\newacronym{\ligo}{LIGO}{Laser Interferometer Gravitational-wave Observatory} 
\newacronym{\Caltech}{Caltech}{California Institute of Technology}
\newacronym{\MIT}{MIT}{Massachusetts Institute of Technology}
\newacronym{\sph}{SPH}{smooth particle hydrodynamics}
\newacronym{\tsi}{TSI}{the Terascale Supernova Initiative}
\newacronym{\wmap}{WMAP}{the Wilkinson Microwave Anisotropy Probe}
\newacronym{\decigo}{DECIGO}{the Deci-Hertz Interferometric Gravitational-wave Observatory} 
\newacronym{\cmbr}{CMBR}{cosmic microwave background}
\newacronym{\ibbh}{IBBH}{intermediate binary black hole}
\newacronym{\bdj}{BDJ}{Brans-Dicke-Jordan}
\newacronym{\bbo}{BBO}{Big Bang Observer}
\newacronym{\decigo}{DECIGO}{Deci-Hertz Gravitational-Wave Observatory}

\def\MPR#1{{\it Moving Puncture Recipe}#1 (MPR#1)\gdef\MPR{MPR}}
\def\ahz#1{apparent horizon#1 (AH#1)\gdef\ahz{AH}}
\def\CM#1{center-of-mass#1 (CM#1)\gdef\CM{CM}}
\def\CLA#1{close-limit approximation#1 (CLA#1)\gdef\CLA{CLA}}
\def\pnw#1{post-Newtonian#1 (PN#1)\gdef\pnw{PN}}
\def\qnm#1{quasi-normal mode#1 (QNM#1)\gdef\qnm{QNM}}
\def\isco#1{innermost stable circular orbit#1 (ISCO#1)\gdef\isco{ISCO}}
\def\eos#1{equation of state#1 (EOS#1)\gdef\eos{EOS}}
\def\tov#1{Tolman-Oppenheimer-Volkoff#1 (TOV#1)\gdef\tov{TOV}}
\def\ns#1{neutron star#1 (NS#1)\gdef\ns{NS}}
\def\bbh#1{binary black hole#1 (BBH#1)\gdef\bbh{BBH}}
\def\bhns#1{black hole -- neutron star#1 (BHNS#1)\gdef\bhns{BHNS}}
\def\nsns#1{neutron star -- neutron star#1 (NSNS#1)\gdef\nsns{NSNS}}
\def\emri#1{extreme mass-ratio inspiral#1 (EMRI#1)\gdef\emri{EMRI}}
\def\emrb#1{extreme mass-ratio binaries#1 (EMRB#1)\gdef\emrb{EMRB}} 
\def\grb#1{gamma-ray burst#1 (GRB#1)\gdef\grb{GRB}}
\def\imbh#1{intermediate mass black hole#1 (IMBH#1)\gdef\imbh{IMBH}}
\def\smbh#1{supermassive black hole#1 (SMBH#1)\gdef\smbh{SMBH}}
\def\bh#1{black hole#1 (BH#1)\gdef\bh{BH}}
\def\ulx#1{ultra-luminous x-ray source#1 (ULX#1)\gdef\ulx{ULX}}
\def\lmxbs{low-mass x-ray Binaries (LMXBs)\gdef\lmxbs{LMXBs}\gdef\lmxb{LMXB}} 
\def\lmxb{low-mass x-ray Binary (LMXB)\gdef\lmxbs{LMXBs}\gdef\lmxb{LMXB}}

\newcommand\apjl{\ref@jnl{ApJ}}%
\newcommand\mnras{\ref@jnl{MNRAS}}%

\begin{document}
\title{Binary Black Holes: Spin Dynamics and Gravitational Recoil}

\author{Frank Herrmann}
\author{Ian Hinder}
\author{Deirdre M. Shoemaker}
\altaffiliation[Also at ]{Department of Physics and Institute for Gravitation and the Cosmos}
\author{Pablo Laguna}
\altaffiliation[Also at ]{Departments of Astronomy \& Astrophysics, Physics and Institute for Gravitation and the Cosmos}
\affiliation{Center for Gravitational Wave Physics\\
The Pennsylvania State University, University Park, PA 16802} 
\author{Richard A. Matzner}
\affiliation{Center for Relativity and Department of Physics\\
The University of Texas at Austin, Austin, TX 78712}

\begin{abstract}
  We present a study of spinning black hole binaries focusing on the
  spin dynamics of the individual black holes as well as on the
  gravitational recoil acquired by the black hole produced by the
  merger.  We consider two series of initial spin orientations away
  from the binary orbital plane. In one of the series, the spins are
  anti-aligned; for the second series, one of the spins points away
  from the binary along the line separating the black holes.  We find
  a remarkable agreement between the spin dynamics predicted at 2nd
  post-Newtonian order and those from numerical relativity.  For each
  configuration, we compute the kick of the final black hole.  We use
  the kick estimates from the series with anti-aligned spins to fit
  the parameters in the \KKF{,} and verify that the recoil along the
  direction of the orbital angular momentum is $\propto \sin\theta$
  and on the orbital plane $\propto \cos\theta$, with $\theta$ the
  angle between the spin directions and the orbital angular momentum.
  We also find that the black hole spins can be well estimated by
  evaluating the isolated horizon spin on spheres of constant
  coordinate radius.
\end{abstract}


\maketitle
\section{Introduction}\label{sec:intro}

Immediately after the discovery of the \MPR{}~\cite{Baker:2005vv,Campanelli:2005dd},  
a recipe providing the ingredients to successfully evolve \bbh{s}, the 
numerical relativity efforts focused on studying the gravitational recoil or kick acquired by 
the \bh{} produced in the merger~\cite{Herrmann:2006ks,Baker:2006vn,Gonzalez:2006md}. 
The main driving force behind these studies has been the astrophysical 
implications of these kicks on the \smbh{s} at the 
centers of galaxies~\cite{Richstone:1998ea,Magorrian:1997hw}.
Specifically, a detailed understanding of these kicks is vital 
to explain the demographics, growth and 
merger rates of \smbh{s}~\cite{Haiman:2004ve,Micic:2006ta}, as well as their 
absence in dwarf galaxies and stellar
clusters~\citep{Madau:2004mq,Merritt:2004xa}.

When viewed in terms of modes of the gravitational radiation emitted by the binary,
kicks arise from the overlap of those modes~\cite{RevModPhys.52.299,2007ApJ...661..430H}. 
A non-vanishing  overlap will be produced if the \bh{s} in the binary have un-equal masses and/or 
are spinning with non-trivial relative orientations. 
For kicks from non-spinning \bbh{s}, the most comprehensive 
numerical relativity study~\cite{Gonzalez:2006md} showed that one can parameterize
the magnitude of the kick velocity as
\begin{equation}
\label{eq:gonzalez}
V = A\,\frac{q^2\,(1-q)}{(1+q)^5}\left[1+B\frac{q}{(1+q)^2}\right]\,,
\end{equation} 
with $A = 1.2\times 10^4\,\KMS$, $B=-0.93$ and $q = M_1/M_2$.
This parameterization was motivated by the scalings originally introduced 
by Fitchett~\cite{Fitchett:1983fc,Fitchett:1984fd}.
From Eq.~(\ref{eq:gonzalez}), the maximum kick has a magnitude of $175\,\KMS$
and occurs at $q = 0.36$ or symmetrized reduced mass
$\eta = M_1\,M_2/M^2 = 0.195$, with $M=M_1+M_2$ the total mass of the binary.
Other mass parameters that will be used are $\delta M \equiv
M_1-M_2$ and $\mu \equiv M_1\,M_2/M$.
When compared to the escape velocities of galactic structures, the kicks from
non-spinning and un-equal mass binaries are modest. 
They are not high enough to eject the \bh{} from its host galaxy~\cite{Merritt:2004xa}. 

The next frontier was to investigate kicks in which the emission of linear 
momentum was due to the spin of the \bh{s}. 
The first study of this kind~\cite{2007ApJ...661..430H} produced 
kick velocities of $V = 475\,\KMS a/M$ for \bh{s} with opposite and equal magnitude spins 
parallel to the orbital angular momentum. Similar studies followed 
soon after~\cite{2007gr.qc.....1163K,2007ApJ...659L...5C} that produced complementary results.
The kick velocities of $\sim 500\,\KMS$ obtained from these configurations 
could in principle explain the absence of massive \bh{s} in dwarf ellipticals~\cite{Merritt:2004xa}.
Motivated by \pnw{} results~\cite{PhysRevD.52.821}, it was immediately realized that
the orientation of the \bh{s'} spins has a profound effect on the kick that
the final \bh{} receives. \citet{Gonzalez:2007hi}
carried out the first simulations in which the spins of the \bh{s} are initially anti-aligned in the orbital plane
and found that kick velocities of at least $2500\,\KMS$ are possible. 
Similar studies~\cite{2007gr.qc.....2133C} suggest that the kick could be scaled to reach 
a maximum of $\sim 4000\, \KMS$. 

As more studies of gravitational recoil continued to emerge, 
generalizations of the phenomenological kick formula Eq.~(\ref{eq:gonzalez}) 
to include spins have been 
introduced~\cite{2007astro.ph..2390B,2007astro.ph..2641S,2007gr.qc.....2133C}, all motivated by the
structure of the formula for the rate of linear momentum radiated, 
a formula first derived by Kidder~\cite{PhysRevD.52.821}.
The terms involving spin-orbit effects in this formula read
\begin{equation}
\label{eq:kidder}
\frac{d\mathbf{P}}{dt}=-\frac{8}{15}\frac{M^3}{r^5}
\frac{q^2}{(1+q)^4} 
\left\{4\dot{r}\left(\mathbf{v}\times \mathbf{\Sigma}\right) 
- 2v^2\left(\mathbf{n}\times \mathbf{\Sigma} \right)-
\left(\mathbf{n}\times \mathbf{v}\right)\left[ 3\dot{r}(n\Sigma)
+2(v\Sigma)\right]\right\}  \,,
\end{equation}
where $(ab)$ denotes the vector dot product, i.e. $(ab)=\mathbf{a}\cdot\mathbf{b}$.
We are following as closely as possible the notation in Ref.~\citep{Faye:2006gx} and
introduce the spin variables
\begin{eqnarray*}
\label{SSigma}
\mathbf{S} &\equiv& \mathbf{S}_1 + \mathbf{S}_2 \\ 
\mathbf{\Sigma}&\equiv& M \Big(\frac{\mathbf{S}_2}{M_2} - \frac{\mathbf{S}_1}{M_1}\Big)\,,
\end{eqnarray*}
where the vector $\mathbf{x}$ denotes the relative position vector of
$M_2$ with respect to $M_1$, with $r=|\mathbf{x}|$, $\mathbf{v} =
d\mathbf{x}/dt$, $\mathbf{n}=\mathbf{x}/r$ and
$\mathbf{L}_\mathrm{N}\equiv\mu\,\mathbf{x}\times\mathbf{v}$, the
Newtonian angular momentum. We also introduce a flat-space orthonormal
rotating triad $\{\mathbf{n},\mathbf{k},\mathbf{l}\}$ such that
$\mathbf{k}=\mathbf{l}\times\mathbf{n}$ with
$\mathbf{l}=\mathbf{L}_\mathrm{N}/\vert\mathbf{L}_\mathrm{N}\vert$
and hence $\mathbf{l}$ is perpendicular to the orbital plane.

With these definitions, Eq.~(\ref{eq:kidder}) has the following structure:
\begin{eqnarray}
\frac{d \mathbf{P}}{d t}  &=& [\dots](\mathbf{k}\times\mathbf{\Sigma}) 
+ [\dots](\mathbf{n}\times\mathbf{\Sigma})
+ \Big\{[\dots](k\Sigma) 
+ [\dots](n\Sigma)\Big\}\mathbf{l}\,,
\end{eqnarray}
or equivalently
\begin{eqnarray}
\label{eq:kidder2}
\frac{d \mathbf{P}}{d t}  &=& 
\Big\{ [\dots]\mathbf{k} + [\dots]\mathbf{n}\Big\} (l\Sigma)
+ \Big\{[\dots](k\Sigma) + [\dots](n\Sigma)\Big\}\mathbf{l}\,,
\end{eqnarray}
where we only show the explicit dependence on $\mathbf{\Sigma}$ relative to
the orthonormal tetrad. Given the form of Eq.~(\ref{eq:kidder2}),
we propose the following parameterization of the contribution of the
spins to the gravitational recoil:
\begin{equation}
\label{eq:kidder-kick}
\mathbf{V}  = \frac{\Sigma}{M^2}\frac{q^2}{(1+q)^4} 
\Big\{ [H_k\mathbf{k} + H_n\mathbf{n}] (l\sigma)
+ [K_k(k\sigma) + K_n(n\sigma)]\mathbf{l}\Big\}\,,
\end{equation} 
where $\sigma = \mathbf{\Sigma}/\vert\mathbf{\Sigma}\vert$. 
We will refer to Eq.~(\ref{eq:kidder-kick}) as the \KKF{.}\footnote{There are several versions
of parameterized kick formulas. Since all are motivated by Kidder's seminal work~\cite{PhysRevD.52.821},
we will generically call them {\it Kidder kick formulae}.}  
The parameters $H_k,\,H_n,\, K_k$ and $K_n$ in Eq.~(\ref{eq:kidder-kick}) are to be determined from
numerical simulations. A fundamental aspect of the validity of this
formula is the dependence of the kick velocity
on the cosine angles $(l\sigma)$, $(k\sigma)$ and $(n\sigma)$.
Spin precession will force these angles to evolve in time.
Thus, one is faced with the task of measuring the \emph{entrance} angles. 
These are the angle values when the binary reaches the ``last'' orbit or plunge, 
namely the time that signals the beginning of the phase when the bulk of the kick gets accumulated.
An identification of the \emph{entrance} angles would 
allow one to determine the $H_k,\,H_n,\, K_k$ and $K_n$ parameters in Eq.~(\ref{eq:kidder-kick}) from
numerical simulations. 

The work in this paper is aimed at exploring the parameter space of
spinning \bbh{s} with focus on the dynamics of the individual spins and 
the kick that the final \bh{} receives. We consider two series of equal mass \bh{s} (i.e. $\delta M=0$).
In one series, called the \emph{B-series}, the \bh{s} initially have equal spin magnitudes and 
anti-aligned directions. That is, $\mathbf{S}=0$ and $\mathbf{\Sigma} = 4\,\mathbf{S}_2 = -4\,\mathbf{S}_1$.
The elements of this series are obtained by changing the orientation of $\mathbf{\Sigma}$ relative to the unit vector $\mathbf{l}$. 
In the second series, called the \emph{S-series}, we also keep the spin magnitudes constant. 
What changes in this series is the relative alignment of the spins.
For each run in both series, we monitor the precession dynamics of the individual spins and compare them with
\pnw{} predictions. We find a remarkable agreement with 2\pnw{} results: the 
2\pnw{} dynamics closely match those from numerical relativity up to the point when a common \ahz{} is formed.
For all models, we compute the gravitational recoil on the final \bh{.} 
We use the kick estimates from the \emph{B-series} to find parameters in the \KKF{} and also
verify the angular dependence in $\mathbf{V}$ that this formula implies.
As numerical relativity efforts explore different regions of the parameter space,
the values of the parameters in Eq.~(\ref{eq:kidder-kick}) will be improved or validated.
A phenomenological formula of this kind is of great value for astrophysical studies 
such as those explaining the population of \smbh{s}.

The paper is organized as follows: In Sec.~\ref{sec:estimates}, we use 
a multipole analysis to demonstrate
the dependence of the kicks on the spin orientations as given
by the \KKF{.} In Sec.~\ref{sec:method}, we summarize our 
computational infrastructure. A detailed description of the two series
of initial data configurations is given in Sec.~\ref{sec:id}. 
The analysis of the \bh{} spin dynamics is presented in Sec.~\ref{sec:spin-dynamics}.
Kick results, including the fit to the \KKF{,} are given in Sec.~\ref{sec:kicks}. 
We end with conclusions in Sec.~\ref{sec:conclusions}.

\section{Kicks and Entrance Angles}
\label{sec:estimates}

To gain further understanding of the \KKF{,} we present an analysis 
based on the multipole formulas of Refs.~\cite{PhysRevD.52.821,RevModPhys.52.299},
in which the rate of radiated linear momentum is estimated, 
to lowest order, as an interference of the mass and spin quadrupoles.
Excluding non-spin terms, this formula reads
\begin{eqnarray}
\label{kick}
\frac{dP^i}{dt} &=& \frac{16}{45} \epsilon^{ijk} I^{(3)}_{jl}
H^{(3)}_{kl}
+ \frac{4}{63} H^{(4)}_{ijk} H^{(3)}_{jk}
+ \frac{1}{126} \epsilon^{ijk} I^{(4)}_{jlm} H^{(4)}_{klm}\,.
\end{eqnarray}
Here $I_{ij}$ and $I_{ijk}$ are respectively the mass quadrupole and octupole.  
Similarly, $H_{ij}$ and $H_{ijk}$ are the spin quadrupole and octupole, respectively.  
In Eq.~(\ref{kick}), a super-index $\,^{(n)}$ denotes an $n$th-time derivative.  

In previous work~\cite{2007ApJ...661..430H}, we used the first term (interference between 
the mass and the spin quadrupoles) to estimate  the kick from 
quasi-circular inspiral to merger by integrating Eq.~(\ref{kick}).
This term is periodic, with period equal to the orbital period, so the kick is dominated by the 
``last'' half orbit in the inspiral. The estimate is computed by 
integrating over a close-in half orbit  (as in Section \ref{sec:intro}, 
the result depends on the magnitude 
and direction of the spins with respect to the orbital angular 
momentum $\mathbf{L} = L \mathbf{l}$), and absorbing the resulting 
error as a normalization constant, where the constant is fixed by 
comparing estimate to numerics for {\it one} configuration. We take the 
same approach here.

Note that the second term in Eq.~(\ref{kick}) will be quadratic in the spin, 
but the spin multipoles have one extra factor $({\bf S}_{1,2}/M_{1,2})/d$ (where 
$d$ is the ``last orbit separation'', and of order several $M$) 
that suppresses the radiation from this term by the same 
factor compared to the first term. While this term's contribution may become 
important in the future, for the moderate spin values we (and others) are 
currently considering, we do not expect significant {\it nonlinear} dependence. 
The third term vanishes (the mass octupole vanishes) 
for equal mass circular orbits as appropriate to our computational quasi-circular 
inspiral, so the equation in our current context is just the first term. 

For the purpose of investigating the \emph{entrance} angles, 
we consider a binary system consisting of equal mass \bh{s} in circular orbit initially confined 
to the $xy$ plane. The orbit is initially oriented so that the \bh{s} are located on the $x$-axis, the
\bh{$_1$} on the positive $x$-axis and \bh{$_2$} on the negative $x$-axis. 
We discuss first the case in which only the \bh{$_1$} is spinning.
We parameterize the orientation of the spin using the usual (fixed frame) polar and axial 
angles $\theta$ and $\varphi$. Thus we have 
$S_{1x}= S_1 \sin\theta\cos\varphi$, $S_{1y}= S_1 \sin\theta\sin\varphi$ and  $S_{1z}= S_1\cos \theta$.

The calculation of the mass quadrupole is straightforward, 
see e.g.~\cite{2007ApJ...661..430H}; the spin quadrupole can be most easily calculated by imagining a spin dipole
(charges $\pm M_1/2$, separation $S_1/M_1$) and conceptually taking the limit
at the end. The spin enters only linearly in the  spin 
quadrupole $H_{kl}$. The structure is different for spin components in 
different directions, and we can compute them independently for the different components.
The nonzero components are:

For $S_{1x}$:  
\begin{eqnarray}
{}^{(x)}H^{(3)}_{xx} &=& \frac{1} {3}\,  d\,S_{1x}\,\omega^3 \sin{(\omega t)}\nonumber\\
{}^{(x)}H^{(3)}_{yx} &=& -\frac{1} {4}\,  d\,S_{1x}\,\omega^3 \cos{(\omega t)}\nonumber\\
{}^{(x)}H^{(3)}_{yy} &=& -\frac{1} {6}\,  d\,S_{1x}\,\omega^3 \sin{(\omega t)}\nonumber\\
{}^{(x)}H^{(3)}_{zz} &=& -\frac{1} {6}\,  d\,S_{1x}\,\omega^3 \sin{(\omega t)}\,;
\label{spinQPx}
\end{eqnarray}

For $S_{1y}$:  
\begin{eqnarray}
{}^{(y)}H^{(3)}_{xx} &=& \frac{1} {6}\,  d\,S_{1y}\,\omega^3 \cos{(\omega t)}\nonumber\\
{}^{(y)}H^{(3)}_{yx} &=&  \frac{1} {4}\,  d\,S_{1y}\,\omega^3 \sin{(\omega t)}\nonumber\\
{}^{(y)}H^{(3)}_{yy} &=& -\frac{1} {3}\,  d\,S_{1y}\,\omega^3 \sin{(\omega t)}\nonumber\\
{}^{(y)}H^{(3)}_{zz} &=&  \frac{1} {6}\,  d\,S_{1y}\,\omega^3 \cos{(\omega t)}\,;
\label{spinQPy}
\end{eqnarray}

For $S_{1z}$:  
\begin{eqnarray}
{}^{(z)}H^{(3)}_{xz} &=& \frac{1} {2}\,  d\,S_{1z}\,\omega^3 \sin{(\omega t)}\nonumber\\
{}^{(z)}H^{(3)}_{yz} &=& -\frac{1} {2}\,  d\,S_{1z}\,\omega^3 \cos{(\omega t)}\,.
\label{spinQPz}
\end{eqnarray}

The spin quadrupole for arbitrary spin direction is the sum of the $S_{1x}, S_{1y}, S_{1z}$ terms. 
In deriving these expressions, we assume that spins, which are parallel transported 
in the evolution, remain constant in Cartesian coordinates. This approximation is 
adequate for the level of accuracy of these estimates. 
The radiated linear momentum equation Eq.~(\ref{kick}) is then explicitly: 
\begin{eqnarray}
\frac{dP^x}{dt} &=& \frac{8}{45} M^2 d^3 \, \omega^6\,S_{1z} \sin{( \omega t)} \nonumber\\
\frac{dP^y}{dt} &=& -\frac{8}{45} M^2 d^3  \, \omega^6 \,S_{1z}\cos{( \omega t)} \,\nonumber\\
\frac{dP^z}{dt} &=& -\frac{16}{45} M^2 d^3 \, \omega^6 [S_{1x} \cos{( \omega t)}- S_{1y} \sin{(\omega t)}] \,.
\label{Pxyz_a}
\end{eqnarray}
The in-plane component of the force rotates with the orbit; the out of plane 
component oscillates at the frequency of the orbit.

If there is a spin on the second hole, the forms are the same as Eqs.~(\ref{Pxyz_a}), 
but the angle $\omega t$ is replaced by $\omega t + \pi$. This replacement has 
the effect of introducing a global minus sign into the spin quadrupole for the
spin on the hole initially located on the negative $x$-axis. 
This means that 
the kick estimate is doubled if the second spin is equal and opposite, but we 
estimate zero kick if the spins are equal and parallel.
For generic second spin, as in our \emph{S-series}, 
one simply subtracts the components in Eqs.~(\ref{Pxyz_a}) 
for this second spin $S_{2}$ from those for the first. 

We concentrate our attention on the \emph{B-series}. As we shall see later, 
this is the series for which we are going to be able to verify, from our simulations,
the dependence of the \KKF{} on the \emph{entrance} angles.
In the \emph{B-series}, the spins are fixed magnitude. Hence, the Eqs.~(\ref{Pxyz_a}), including
the contributions from both spins, read:
\begin{eqnarray}
\label{eq:richardFormula}
\frac{dP^x}{dt} &=&  \frac{16}{45} M^2 d^3 \, \omega^6 \,S_1 \cos\theta\sin{( \omega t)} \nonumber\\
\frac{dP^y}{dt} &=& -\frac{16}{45} M^2 d^3 \, \omega^6 \,S_1 \cos\theta\cos{( \omega t)} \,\nonumber\\
\frac{dP^z}{dt} &=& -\frac{32}{45} M^2 d^3 \, \omega^6 \,S_1 \sin\theta\cos{( \omega t + \varphi)} \,.
\label{Pxyz}
\end{eqnarray}
Eqs.~(\ref{eq:richardFormula}) predict a $z$-kick $V^z \propto \sin \theta$ and
kicks $\propto \cos \theta$ in the orbital plane. Notice also the dependence of the 
$z$-kick on the entry angle $( \omega t + \varphi)$, demonstrating the fact 
that the net $z$-kick can vanish for carefully chosen entry angle. 
For the circular orbits treated here, $dP^z/dt$ in Eqs.~(\ref{eq:richardFormula}) 
identifies the quantities $K_k$ and $K_n$ in the \KKF{}, Eq.~(\ref{eq:kidder-kick}), as equal.
We will compare the predictions of Eqs.~(\ref{eq:richardFormula}) on the scaling of the
kicks with the angle $\theta$ in Sec.~\ref{sec:kicks}.

\section{Computational Methodology}\label{sec:method}
We follow the \MPR{} to evolve the \bbh{} configurations. Briefly, the 
\MPR{} builds upon the BSSN system of evolution equations~\cite{Nakamura87,
  Shibata95,Baumgarte99}, models \bh{s} with ``punctures''~\cite{Brandt97b}
and uses dynamic gauge conditions~\cite{Baker:2005vv,Campanelli:2005dd} designed to allow
these punctures to move. The explicit form of the evolution equations
for the lapse and shift gauge quantities are the ``covariant'' form of the ``1+log'' slicing~\cite{Bona97a} $(\partial_t -
\beta^i\partial_i)\alpha = - 2\alpha K$ and a modified gamma-freezing condition~\cite{Alcubierre02a,vanMeter:2006vi} for the shift: $\partial_t \beta^i = B^i, B^i=\partial_t \widetilde{\Gamma}^i -\xi \partial_t\beta^i -\beta^j\partial_j \widetilde{\Gamma}^i$, where $K$
is the trace of the extrinsic curvature, $\widetilde\Gamma^i$ the
trace of the conformal connection and $\xi=2$ a free, dissipative
parameter. The importance of these gauge conditions is twofold: First,
they avoid the need of excising the \bh{} singularity from the
computational domain since they effectively halt the evolution (i.e.
lapse function $\alpha$ vanishes) near the \bh{}
singularity~\cite{Hannam:2006vv}. Second, they allow for movement of
the \bh{} or {\it puncture} through the computational domain while
freezing the evolution inside of the \bh{} horizon. See Ref.~\cite{2006gr.qc....10128B} for a detailed description and analysis of the \MPR{}.

Our source code was produced by the \texttt{Kranc} code
generation package~\cite{Husa:2004ip} and uses the \texttt{Cactus}
infrastructure~\cite{Cactusweb} for parallelization and
\texttt{Carpet}~\cite{Schnetter-etal-03b} for mesh refinement.  The
code uses fourth order accurate finite differencing (centered for all non-advection and a lop-sided stencil for the advection terms) and a fourth order Runge-Kutta temporal updating scheme
with Courant factor of 0.5.
The initial data code was developed by~\citet{Ansorg:2004ds}. The
initial free parameters (e.g. specifying angular momentum, spins,
masses, separations) are chosen according to the effective
potential method~\cite{Cook94,Baumgarte00a} or using \pnw{}
parameters~\cite{Campanelli:2006uy,Gonzalez:2007hi}. These methods
both yield \bbh{} initial data sets representing \bbh{s} in
quasi-circular orbit~\cite{2006gr.qc....10128B}.

The computational grids consist of a nested set of 10 refinement
levels, with the finest mesh having resolution $h=M/35.2$. This
resolution translates into a resolution of about $h=m/14$, with
respect to the bare mass, $m$, of the punctures according to
Tables~\ref{tbl:ID-B-series} and \ref{tbl:ID-S-series}. The minimal resolution
found to be adequate for spinning cases according
to~\citet{Campanelli:2006uy} is $h < M/30$.  The grid sizes in our
$h=M/35.2$ simulations are: the 4 finest refinement levels have $44^3$ grid-points 
plus 6 coarser refinement levels of $88^3$.  All grids are initially cubical. During
the evolution, the shape and number of grid-points per refinement level
vary due to adaptivity. The coarsest mesh is kept fixed and extends to
$640\,M$ from the origin in each direction. Because the simulations in this work are very similar
(regarding mesh setups, grid sizes and refinement scales) 
to those in our previous work~\cite{2007ApJ...661..430H}, the convergence and
errors estimates in the present study are comparable.

In order to study the spin dynamics of the \bh{s}, we need
infrastructure to compute the individual spins of the \bh{s}.  The
isolated horizon
formalism~\cite{Dreyer02a,Ashtekar:2004cn,Schnetter:2006yt} provides a
definition associated with a Killing vector of the spacetime of the
spin of a single \bh{}:
\begin{equation}\label{eq:IH-spin}
  S_\varphi=\frac{1}{8\pi}\oint_{AH} \varphi^i n^j K_{ij} dS
\end{equation}
where $\varphi^i$ is a Killing vector on the \ahz{} surface, 
$K_{ij}$ is the extrinsic curvature of the 3D-slice and $n^i$ is the outward pointing unit normal
vector to the \ahz{.} The direction of the spin is given by
the Killing vector $\varphi^i$. To facilitate finding the spin direction, 
\citet{2006gr.qc....12076C} introduced the
usage of the flat space coordinate rotational Killing vectors 
\begin{eqnarray*}
\varphi_x^i &=& (0,-\hat z,\hat y)\\
\varphi_y^i &=& (\hat z,0,-\hat x)\\
\varphi_z^i &=& (-\hat y,\hat x,0)\,,
\end{eqnarray*} 
where the coordinates $(\hat x,\hat y,\hat z)$ are relative to the position of the \bh{.}
The spin is then given by $\mathbf{S} = (S_x,S_y,S_z)$, where each component is obtained, in the
fixed $\lbrace x,y,z \rbrace$ coordinate system, by
evaluating Eq.~(\ref{eq:IH-spin}) with each of the coordinate rotational
Killing vectors. There is an excellent agreement between the approximate spin 
this method yields and the one using 
the Killing vector $\varphi^i$ (when one exists)~\cite{2006gr.qc....12076C}. 
There are efficient \ahz{} finders~\cite{Thornburg2003:AH-finding_nourl} available; however, they impose a
non-negligible overhead in the simulations. To gain efficiency, we relax the condition that the
integral in Eq.~(\ref{eq:IH-spin}) has to be evaluated at the \ahz{} and
choose a coordinate sphere around the puncture. The radius of the sphere is chosen sufficiently small,
that the sphere is contained within the \bh{'s} horizon.

\begin{figure}
\begin{center}
  \includegraphics{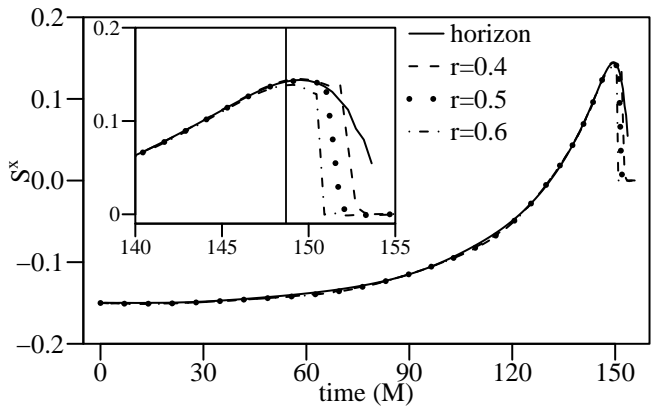}
\end{center}
\caption{Comparison of $S^x$ computed on the horizon and from spheres
  with radius $r$ for a \bbh{} evolution (S-90 model, see Table~\ref{tbl:ID-S-series}). The vertical line (here and in subsequent Figures)
shows the first time a common \ahz{} is
found.}
\label{fig:ihspin_cmp}
\end{figure}

Fig.~\ref{fig:ihspin_cmp} shows a comparison of the $S_x$ component between
the values using the \ahz{} surface and three different
coordinate spheres of radius $r$ for the S-90 model (see Table~\ref{tbl:ID-S-series}) \bbh{} evolution. There is good agreement into the merger regime.
The vertical line in Fig.~\ref{fig:ihspin_cmp} and subsequent figures shows the first time a common \ahz{} is
found. After that time, no individual apparent horizons exist 
and the spheres centered on the punctures track different and meaningless values of $S_x$.

\section{Initial Data and Radiated Quantities}\label{sec:id}
We consider two series of equal mass \bh{s} (i.e. $\delta M=0$).
In both series, initially the \bh{s} have the same spin magnitude $S_1/M_1^2 = S_2/M_2^2 =0.6$.
The initial orientation of the \bh{'s} spin is in the 
$xz$-plane. We use the polar angle $\theta_1$: the angle 
between the $z$-axis and the direction of $\mathbf{S}_1$ in the $xz$-plane.
The \bh{$_2$} is treated similarly. 
We take the convention that positive (negative) $\theta$ angles are measured 
(counter-) clockwise in the $xz$-plane.
In the series referred to as \emph{B-series},
the \bh{'s} are anti-aligned, i.e. $\theta \equiv \theta_2 = \theta_1 - 180^o$, so
$\mathbf{S}_1=-\mathbf{S}_2$. That is, $\mathbf{S}=0$ and $\mathbf{\Sigma} = 4\,\mathbf{S}_2 = -4\,\mathbf{S}_1$.
The elements in this series are obtained by changing $\theta$. 
In the \emph{S-series}, we initially orient ${\bf S_1}$ to
$\theta_1 = 270^o = -90^o$ and vary $\theta \equiv \theta_2$.

We chose orbital parameters (i.e. bare masses, separation and momentum)
in the \emph{B-series} by minimizing the effective binding
energy~\cite{1994PhRvD..50.5025C,Baumgarte00a}, while for the \emph{S-series}
we used \pnw{} parameters~\cite{Campanelli:2006uy,Gonzalez:2007hi}.
Initially, \bh{$_1$} is located at position $(-x/M,0,0)$ and has 
linear momentum $(0,-p_y/M,0)$.
Similarly, \bh{$_2$} is at position $(x/M,0,0)$ with linear momentum $(0,p_y/M,0)$ . 
It turns out that the bare puncture masses for both series are
roughly constant, $m_1 = m_2 \approx 0.395\,M$ 
to the 3rd digit of precision. The slight changes are needed to keep the
irreducible masses $M_1 = M_2 = 0.5\,M$.
As mentioned above, the spins in both \bh{s} are initially in  the $xz$-plane;
that is, $(S^x_{1,2}/M^2,0,S^z_{1,2}/M^2)$, where 
$S_{1,2}^x = S_{1,2}\,\sin{\theta_{1,2}}$ and $S_{1,2}^z = S_{1,2}\,\cos{\theta_{1,2}}$ with
$S_{1,2} = 0.15\,M^2$.

Table~\ref{tbl:ID-B-series} lists the relevant initial data parameters
for the \emph{B-series}, while Table~\ref{tbl:ID-S-series} gives the
parameters for the \emph{S-series}.
In addition to the initial data parameters, 
the tables also report the radiated angular momentum
$J^z_{\mathrm{rad}}$ in \% of the initial orbital angular momentum,
$L^z_o$, as well as a time estimate of the common AH formation. We use the
maximum in $\Psi_4$ shifted by the extraction radius and an additional $10\,M$ as an indicator for the
merger time $T_\mathrm{max}$. We have found that this measure is accurate
to a few $M$.
For the \emph{B-series}, the spin of the final \bh{} is
$J/M^2=0.62$ for all models. Constant in both series
is the total ADM mass, $E_{\mathrm{ADM}}\approx 0.985\,M$.  
While the runs B-90 and S-90 have the
same spin configurations, i.e. spins pointing along the $x$-axis only,
the radiated energy and angular momentum are different because they
differ in initial separation and angular momentum. 
The radiated quantities were extracted at $r=40\,M$. 
For a number of models, we have carried out simulations at lower resolution ($M/32$) and 
measured at detector radii $r/M=\lbrace 30,40,50,60,80\rbrace$. 
Based on the variations observed in the measured quantities (energy, angular momentum and kicks),
we estimate the reported numbers to be accurate to about 15\%.
\begin{table}
  \begin{center}
\begin{tabular}{ccccccccc}
model & $x [M]$ & $p_y [M]$ & $V [\KMS]$ & $J_\mathrm{rad} [\%\,L^z_o]$ & $E_{\mathrm{rad}}[\%\,M]$& $T_\mathrm{max}[M]$ \\
\hline 
B-20 & 2.986 & 0.138 & 427 & 24 & 3.3 & 109.1\\ 
B-30 & 2.990 & 0.138 & 544 & 24 & 3.3 & 109.1\\ 
B-50 & 3.000 & 0.137 & 761 & 25 & 3.4 & 108.6\\ 
B-70 & 3.009 & 0.137 & 908 & 25 & 3.4 & 108.6\\ 
B-80 & 3.012 & 0.137 & 945 & 25 & 3.4 & 108.4\\ 
B-90 & 3.013 & 0.137 & 963 & 25 & 3.4 & 108.4\\ 
  \hline 
  \end{tabular}
  \end{center}
  \caption{\emph{B-series}: Initial data parameters for the \emph{B-series}. The models in this series
are labeled as B-$\theta$, where the angle $\theta \equiv \theta_2 = \theta_1 - 180^o$
($\theta = 0^o$ corresponds to spins parallel and anti-parallel to the orbital angular momentum).
The punctures have bare masses $m_{1,2}=0.395$, are located on the $x$-axis at $\mp x$ 
and have initial momentum $\mp p_y$ in the $y$-direction.
Results listed are the magnitude of the recoil velocity $V$, 
the radiated angular momentum $J^z_{\mathrm{rad}}$ in \% of the initial orbital angular momentum $L^z_o$, 
the energy radiated $E_{\mathrm{rad}}$, 
and the time $T_\mathrm{max}$ which is an estimate of the merger time derived from the time it takes in each simulation to reach the maximum amplitude in $\Psi_4$.}
\label{tbl:ID-B-series}
\end{table}
\begin{table}
  \begin{center}
\begin{tabular}{cccccccccc}
model & $m_{1,2}[M]$ & $p_y [M]$ & $V[\KMS]$ & $J_\mathrm{rad}[\%\,L^z_o]$ & $E_{\mathrm{rad}}[\% \,M]$& $J^z_{\mathrm{final}}[M^2]$ & $T_\mathrm{max}[M]$\\
\hline 
S-0 & 0.396 & 0.132 & 854 & 34 & 4.6 & 0.68 & 192.3 \\ 
S-15 & 0.396 & 0.132 & 1401 & 33 & 4.4 & 0.68 & 189.5 \\ 
S-30 & 0.396 & 0.132 & 2000 & 33 & 4.4 & 0.67 & 184.1 \\ 
S-45 & 0.396 & 0.133 & 2030 & 32 & 4.3 & 0.66 & 177.3 \\ 
S-60 & 0.395 & 0.134 & 1218 & 30 & 4.0 & 0.65 & 168.6 \\ 
S-75 & 0.395 & 0.135 & 230 & 28 & 3.7 & 0.64 & 159.1 \\ 
S-90 & 0.395 & 0.137 & 1462 & 26 & 3.4 & 0.62 & 148.6 \\ 
S-105 & 0.395 & 0.138 & 1979 & 25 & 3.3 & 0.60 & 138.6 \\ 
S-120 & 0.395 & 0.139 & 1787 & 24 & 3.2 & 0.58 & 130.5 \\ 
S-135 & 0.395 & 0.140 & 1234 & 23 & 3.0 & 0.56 & 124.1 \\ 
S-150 & 0.395 & 0.141 & 689 & 21 & 2.9 & 0.55 & 119.5 \\ 
S-165 & 0.395 & 0.141 & 335 & 21 & 2.8 & 0.55 & 117.7 \\ 
S-180 & 0.395 & 0.141 & 188 & 20 & 2.8 & 0.55 & 117.7 \\ 
S-195 & 0.395 & 0.141 & 157 & 20 & 2.8 & 0.55 & 120.5 \\ 
S-210 & 0.395 & 0.141 & 173 & 22 & 3.0 & 0.56 & 125.5 \\ 
S-225 & 0.395 & 0.140 & 223 & 22 & 3.2 & 0.57 & 132.7 \\ 
S-240 & 0.395 & 0.139 & 268 & 23 & 3.4 & 0.59 & 141.4 \\ 
S-285 & 0.395 & 0.135 & 253 & 26 & 3.9 & 0.65 & 174.1 \\ 
S-300 & 0.396 & 0.134 & 406 & 29 & 4.2 & 0.66 & 181.8 \\ 
S-315 & 0.396 & 0.133 & 399 & 31 & 4.5 & 0.67 & 187.7 \\ 
S-330 & 0.396 & 0.132 & 354 & 32 & 4.6 & 0.68 & 191.8 \\ 
S-345 & 0.396 & 0.132 & 459 & 33 & 4.6 & 0.68 & 193.2 \\ 
  \hline 
  \end{tabular}
  \end{center}
  \caption{The \emph{S-series}. For all cases, initially the \bh{$_1$} 
is located along the $x$-axis at $x=-3.1\,M$,
has momentum pointing along the $y$-direction with value $-p_y$,
and has spin $\mathbf{S}_1=(-0.15\,/M^2,0,0)$, thus $\theta_1 = -90^o$
and $\varphi_1 = -180^o$. 
\bh{$_2$} is located also along  the $x$-axis but at $x = 3.1\,M$ with
momentum $p_y$. 
In these runs, labeled S-$\theta$, the angle $\theta$ gives the
angle in the $xz$-plane that the spin of \bh{$_2$}
makes with respect to the $z$-axis.
Results listed are the magnitude of the recoil velocity $V$, 
the radiated angular momentum $J^z_{\mathrm{rad}}$ in \% of the initial orbital angular momentum $L^z_o$,
the energy radiated $E_{\mathrm{rad}}$,  
the spin of the final \bh{} $J^z_{\mathrm{final}}$ along the $z$-axis, 
and the time $T_\mathrm{max}$ which is an estimate of the merger time derived from the time it takes in each simulation to reach the maximum amplitude in $\Psi_4$.}
\label{tbl:ID-S-series}
\end{table}
%
\section{Spin Dynamics}\label{sec:spin-dynamics}

In the present work, we are 
interested investigating the degree to which the 
spin dynamics described by \pnw{} equations agrees with that from numerical relativity.
Following Ref.~\cite{2006PhRvD..74j4034B}, the precession equation of 
\bh{$_1$} in the binary with mass $M_1$, spin $\mathbf{S}_1$, position $\mathbf{x}_1$ and
velocity $\mathbf{v}_1$ is given by
\begin{equation}\label{eq:precession}
  \frac{d \mathbf{S}_1}{d t} = \mathbf{\Omega}_1\times
  \mathbf{S}_1\,;
\end{equation}
which implies that \bh{$_1$} precesses around the vector $\mathbf{\Omega_1}$ with rate
$\vert\mathbf{\Omega}_1\vert$. The precession angular frequency vector $\mathbf{\Omega_1}$ is given to 
2\pnw{} by
\begin{align}\label{eq:omega}
  \mathbf{\Omega}_1 &= \frac{M_2}{r^2} \bigg[ \frac{3}{2}
  \mathbf{n}_{12}\times \mathbf{v}_1 - 2 \mathbf{n}_{12}\times
  \mathbf{v}_2\bigg] \nonumber \\ & + \frac{M_2}{r^2}
  \bigg[ \mathbf{n}_{12}\times \mathbf{v}_1 \Big(-\frac{9}{4}
  (n_{12}v_2)^2 + \frac{1}{8} v_1^2 - (v_1v_2) + v_2^2 + \frac{7}{2}
  \frac{M_1}{r} - \frac{1}{2} \frac{M_2}{r} \Big)
  \nonumber \\ & \qquad\quad + \mathbf{n}_{12}\times \mathbf{v}_2
  \Big(3 (n_{12}v_2)^2 + 2 (v_1v_2) - 2 v_2^2 + \frac{M_1}{r} +
  \frac{9}{2} \frac{M_2}{r} \Big) \nonumber \\ & \qquad\quad +
  \mathbf{v}_1 \times \mathbf{v}_2 \Big( 3 (n_{12}v_1) - \frac{7}{2}
  (n_{12}v_2) \Big)\bigg]\,,
\end{align}
with $\mathbf{x}=\mathbf{x}_1-\mathbf{x}_2$,
$r=|\mathbf{x}|$ and $\mathbf{n}_{12}=\mathbf{x}/r$.
The expressions for the companion \bh{$_2$} are obtained by 
switching $1\leftrightarrow 2$ in
Eqs.~(\ref{eq:precession}-\ref{eq:omega}).  In Eq.~(\ref{eq:omega}), the
first term in square brackets represents the 1\pnw{} contribution.
For comparison, we also show the precession angular frequency from
Kidder~\cite{PhysRevD.52.821}, in which the terms $\propto {\bf L}_N$ 
(corresponding to the first line in Eq.~(\ref{eq:omega})) are accurate to 1\pnw{} but the expression also contains 
spin-spin terms:
\begin{equation}\label{eq:kidder-prec}
\mathbf{\Omega}_1=\frac{1}{r^3}\bigg[ \mathbf{L}_N \left(2+\frac{3}{2}\frac{M_2}{M_1} \right) 
    - \mathbf{S}_2 + 3(n_{12}S_2)\mathbf{n}_{12}\bigg]\ ,
\end{equation}
where $\mathbf{L}_N =\mu \mathbf{x}\times\mathbf{v}_{12}$ 
denotes the Newtonian angular momentum. Here again one obtains the expression for 
\bh{$_2$} by switching $1\leftrightarrow 2$. 

\begin{figure}
\begin{center}
\includegraphics{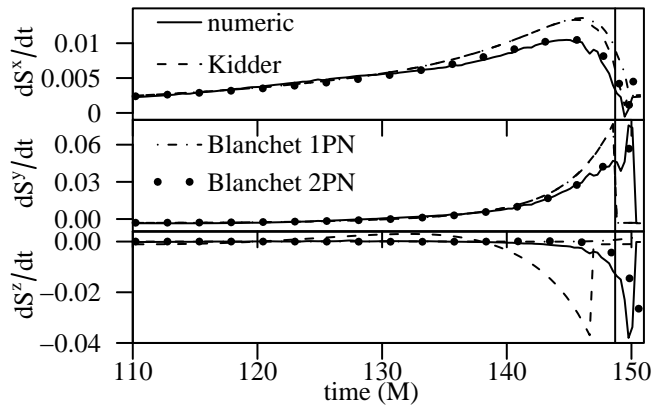}
\end{center}
\caption{Comparison of $d\mathbf{S}_1/dt$ computed from the numerical evolution
  directly and by using \pnw{} formulas for the S-90 run. {\it Kidder} describes the dynamics using
  precession angular frequency given by Eq.~(\ref{eq:kidder-prec}).
  {\it Blanchet 1PN} denotes the dynamics with $\mathbf{\Omega}_{1}$
  given by the first term in Eq.~(\ref{eq:omega}); {\it Blanchet 2PN}
  denotes the case in which the entire expression in Eq.~(\ref{eq:omega}) is used.
The vertical line around $t=149M$ indicates the formation of a common apparent horizon.}
\label{fig:compare_PN_num}
\end{figure}

Fig.~\ref{fig:compare_PN_num} shows the time evolution of $d\mathbf{S}_{1}/dt$
computed in four different ways for the S-90 run (the spins are equal 
magnitude and anti-aligned in the $xy$-plane).
The time evolution of $d\mathbf{S}_{2}/dt$ is equal and opposite in this case.  
Solid lines, labeled \emph{numeric}, represent the numerical relativity solutions.
The values of $d\mathbf{S}_{1}/dt$ are obtained 
by constructing each \bh{} spin as described in Sec.~\ref{sec:method}, 
followed by finite differences to approximate the time derivative.
A long dashed line, labeled \emph{Kidder}, denotes $d\mathbf{S}_{1}/dt$ computed 
using the precession angular frequency Eq.~(\ref{eq:kidder-prec}).
The dotted line, labeled \emph{Blanchet 1\pnw{}}, represents the 
result from using only the 1\pnw{} contribution in the precession
angular frequency Eq.~(\ref{eq:omega}); that is, it corresponds to Kidder's precession
without the inclusion of spin-spin interactions. Finally,
the dashed-dotted line, labeled \emph{Blanchet 2\pnw{}}, depicts the
evolution of  $d\mathbf{S}_{1}/dt$ using the entire expression in Eq.~(\ref{eq:omega}).
In the construction of the \pnw{} precession angular frequencies, 
we use the positions and velocities of the punctures from the 
numerical simulations. The vertical lines in Fig.~\ref{fig:compare_PN_num}
denote the time at which a common \ahz{} is formed. 

It is remarkable how accurately the 2\pnw{} approximations of 
$d\mathbf{S}_{1}/dt$ track the numerical result deep into the merger regime, close to the
formation of a common \ahz{.}
Comparisons beyond the time when a common \ahz{} forms are not very meaningful since the 
individual trapped surfaces loose their horizon interpretation and
our spin measure breaks down (see Sec.~\ref{sec:method}).
Also interesting is that the spin-spin terms in Kidder's expression make only a small
contribution to $dS^x/dt$ and $dS^y/dt$, as can be seen from the similarities 
of the Kidder and Blanchet 1\pnw{} lines. 
On the other hand, the spin-spin are responsible for the differences
between the Kidder and Blanchet 1\pnw{} values of $dS^z/dt$ near the mergers, as one can 
observe in the bottom panel of Fig.~\ref{fig:compare_PN_num}. This discrepancy can be traced to 
the $z$-component in the third term in Eq.~(\ref{eq:kidder-prec}). 
The first term when using the frequency Eq.~(\ref{eq:kidder-prec}) in
Eq.~(\ref{eq:precession}) contains the $z$-component of $\mathbf{L}_N\times \mathbf{S}_1$, 
which is numerically very close to zero for the S-90 model. The second term contains the $z$-component 
of $\mathbf{S}_2\times \mathbf{S}_1$, which is also close to zero. In the third term, 
we have $(n_{12}S_2)$ and the $z$-component of $\mathbf{n}_{12}\times \mathbf{S}_1$. 
Both of these terms grow rapidly near the merger. In particular, the $z$-component 
of $\mathbf{n}_{12}\times \mathbf{S}_1$ develops significant noise which terminates the line early. 
Finally, it is very clear that including terms up to 2\pnw{} makes an important difference in improving
the matching to the numerical solution.

\begin{figure}
\begin{center}
\includegraphics{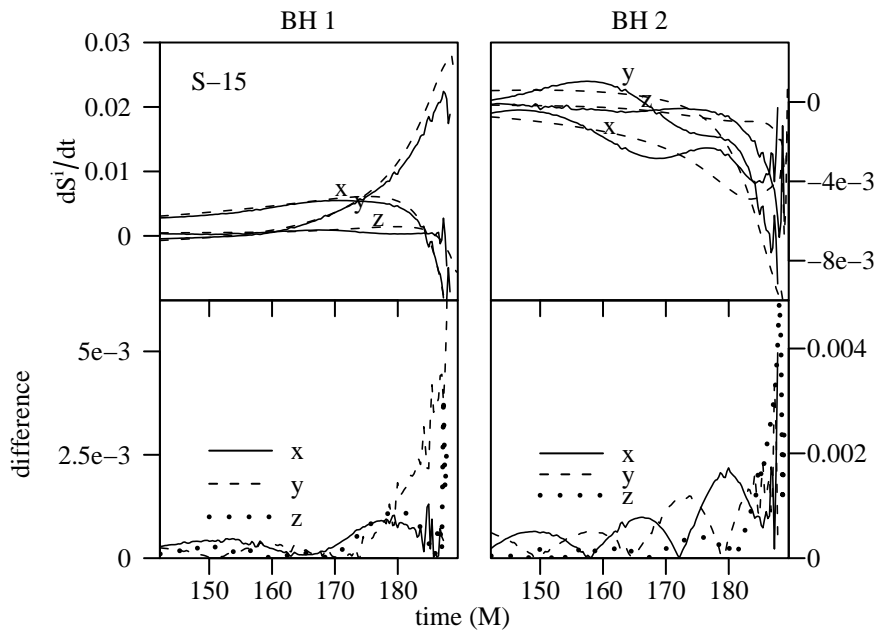}
\end{center}
\caption{Comparison of S-15 run numerical to Blanchet 2PN. Left panel shows the 
results of the comparison for \bh{$_1$} and the right panel for \bh{$_2$}.
The top plots on each panel show with a solid line $dS^i/dt$ from our numerical
simulations and with a dashed line the values from Blanchet 2PN. The labels denote
each component. The bottom plots on each panel show the difference between the
numerical solution and the Blanchet 2PN, with solid, dashed and dotted lines for the
$x$, $y$ and $z$ components, respectively.}
\label{fig:S-15_overview}
\end{figure}
\begin{figure}
\begin{center}
\includegraphics{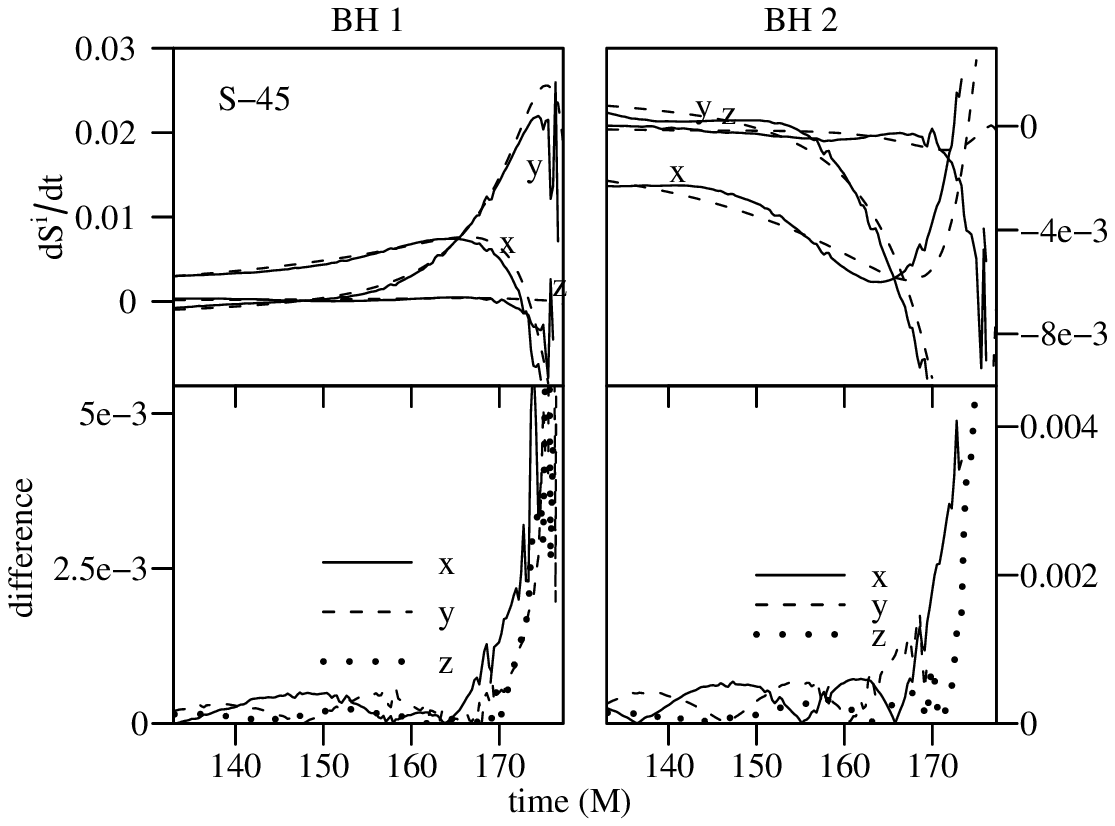}
\end{center}
\caption{Same comparison as in Fig.~\ref{fig:S-15_overview} but for the model S-45.}
\label{fig:S-45_overview}
\end{figure}
\begin{figure}
\begin{center}
\includegraphics{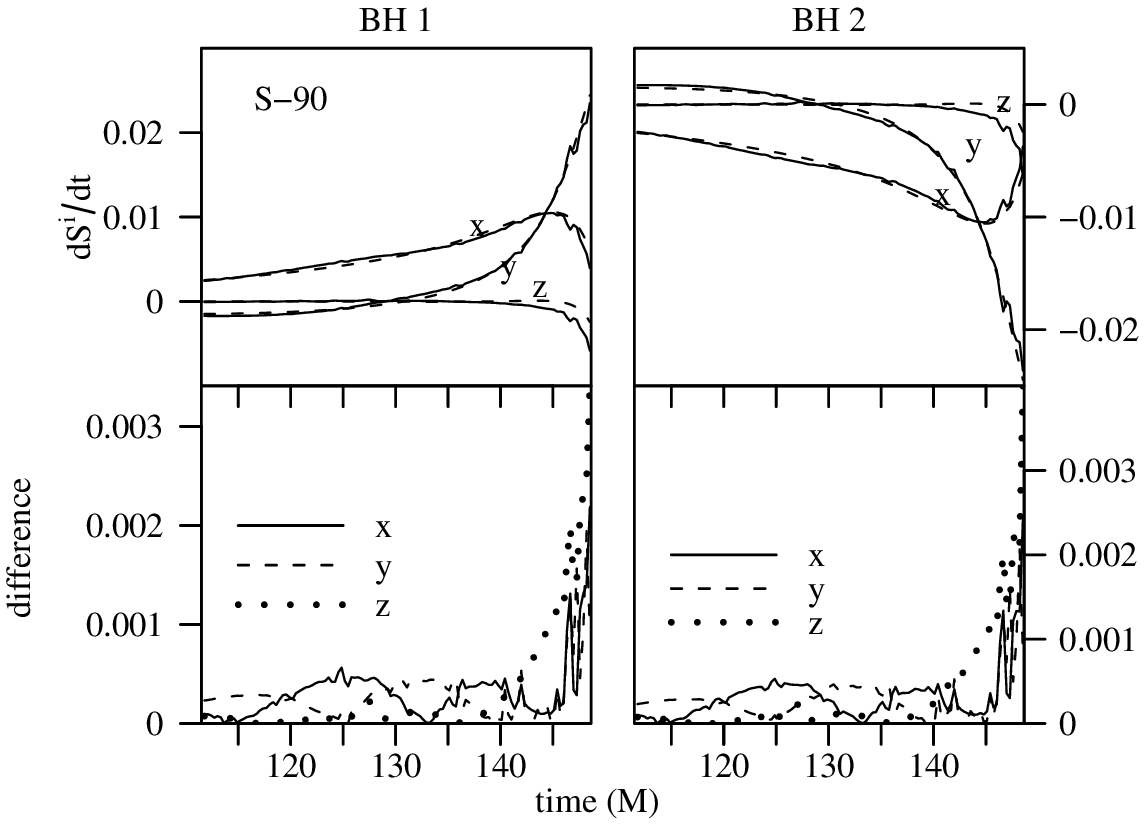}
\end{center}
\caption{Same comparison as in Fig.~\ref{fig:S-15_overview} but for the model S-90.}
\label{fig:S-90_overview}
\end{figure}
\begin{figure}
\begin{center}
\includegraphics{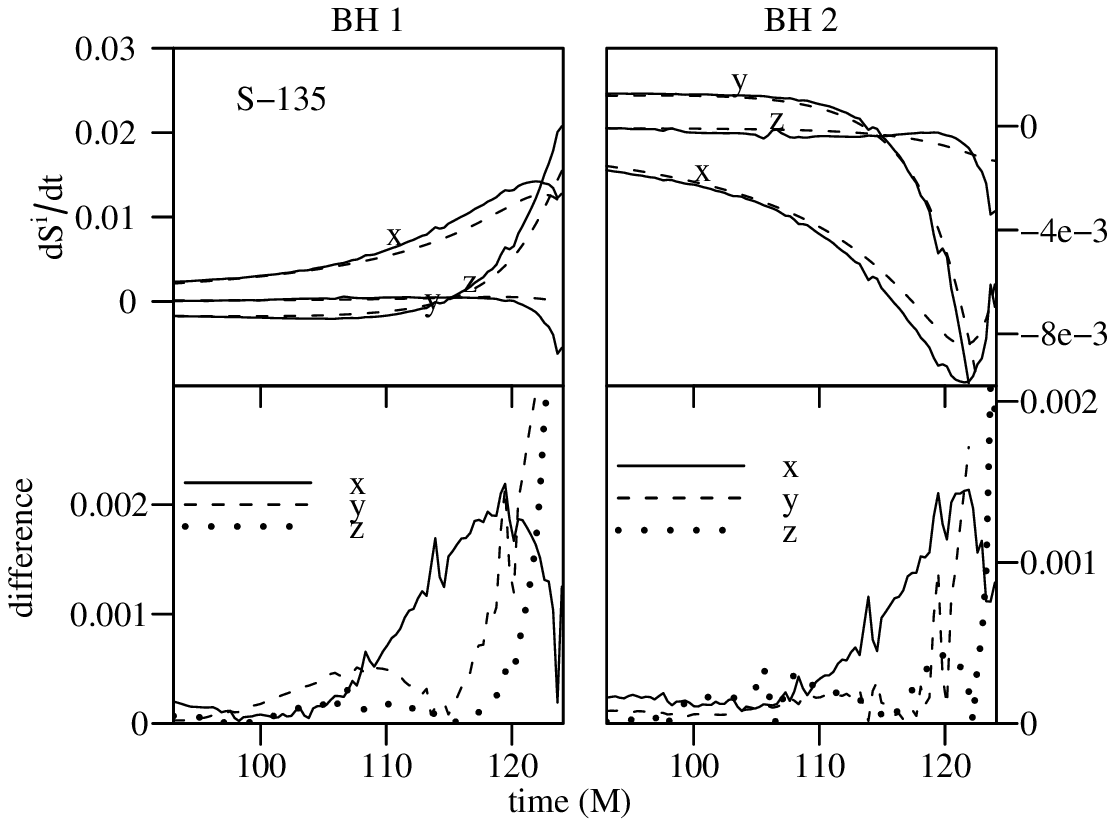}
\end{center}
\caption{Same comparison as in Fig.~\ref{fig:S-15_overview} but for the model S-135.}
\label{fig:S-135_overview}
\end{figure}
\begin{figure}
\begin{center}
\includegraphics{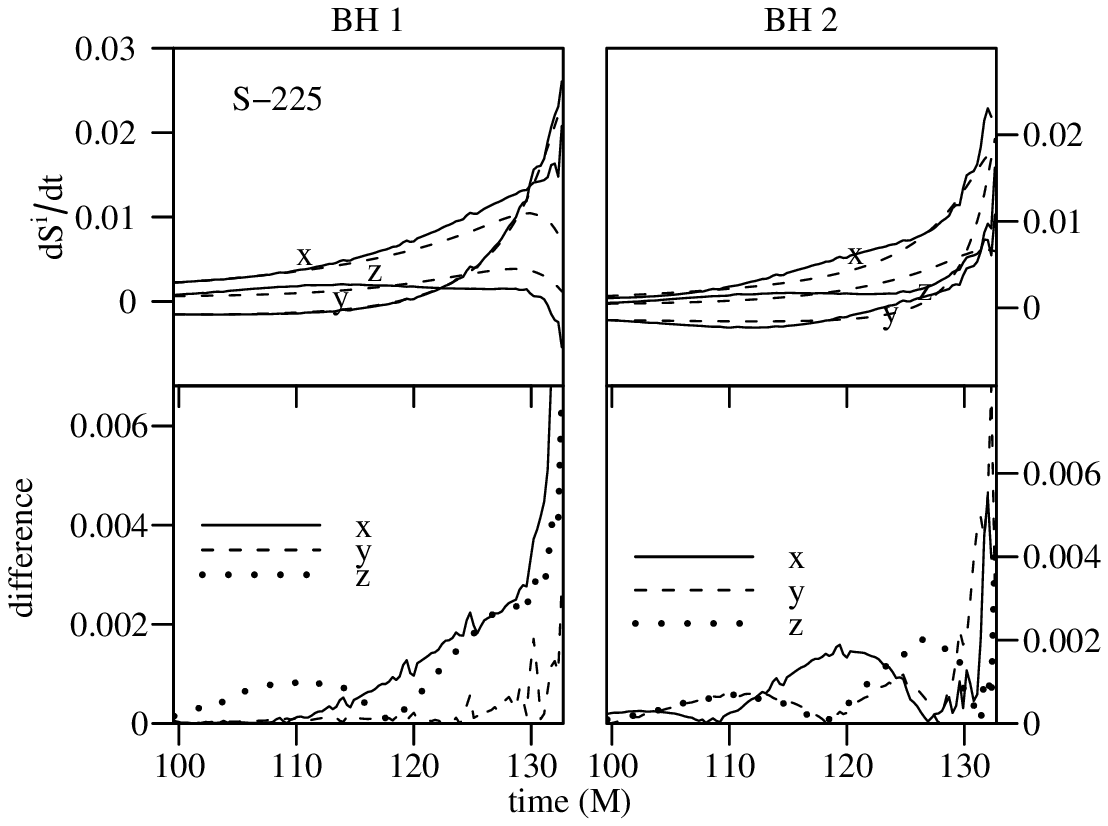}
\end{center}
\caption{Same comparison as in Fig.~\ref{fig:S-15_overview} but for the model S-225.}
\label{fig:S-225_overview}
\end{figure}
We have carried out comparisons similar to that in
Fig.~\ref{fig:compare_PN_num} for all the runs in the \emph{B-} and
\emph{S-series}.  The results in every case are similar; namely, that the dynamics
of $d\mathbf{S}_{1,2}/dt$ are very well approximated by 2\pnw{,} and this description only
starts breaking down close to the merger.

The most significant variation observed in the $d\mathbf{S}_{1,2}/dt$
dynamics from one S-series run to another was the time at which the
merger takes place or, equivalently, the time at which the
gravitational radiation emitted reaches its maximum (see Table~
\ref{tbl:ID-S-series}).  The differences in merger time are due to
spin hangup~\cite{2006gr.qc....12076C} of the merger.
Figs.~\ref{fig:S-15_overview}-\ref{fig:S-225_overview} show the
comparison for a few selected models from the \emph{S-series}. The
plots show $d\mathbf{S}/dt$ for both \bh{s}. The left panel shows
\bh{$_1$}, which has the initial spin direction along the negative
$x$-axis. In this case, there is good agreement between the 2PN and
computational results for all models. In the right panel, we show
\bh{$_2$} with a different initial spin direction specified according
to the angle $\theta$.  For spins of \bh{$_2$} more parallel to the
orbital angular momentum, the precession becomes smaller (note range
of the $y$-axis to the right of the figure). It seems very likely that
with the small precession shown in the S-15 model, the visible
disagreement to the \pnw{} result is just a numerical artifact that
could be cured by higher resolution.
\begin{figure}
\begin{center}
  \includegraphics{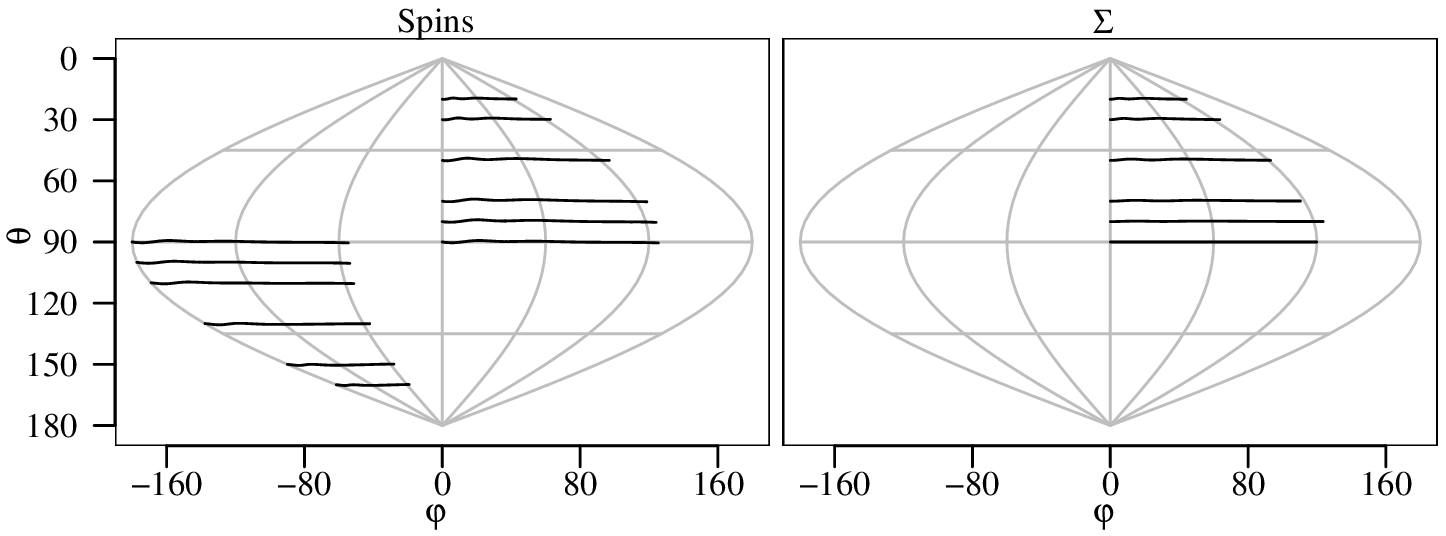}
\end{center}
\caption{The evolution tracks of the $\mathbf{S}_{1,2}$ and $\mathbf{\Sigma}$ directions in
  the $\theta$-$\varphi$ plane for all the cases in the \emph{B-series}.
  The left plot shows the individual spins and the right plot shows
  the evolution of $\mathbf{\Sigma}$.
  All the cases start with $\varphi_{1}=-180^o$ and $\varphi_{2}=0^o$. Notice that
  there is almost no change in the $\theta_{1,2}$ direction for the
  individual spins or for $\mathbf{\Sigma}$.}
\label{fig:sky_map_b_series}
\end{figure}

\begin{figure}
\begin{center}
  \includegraphics{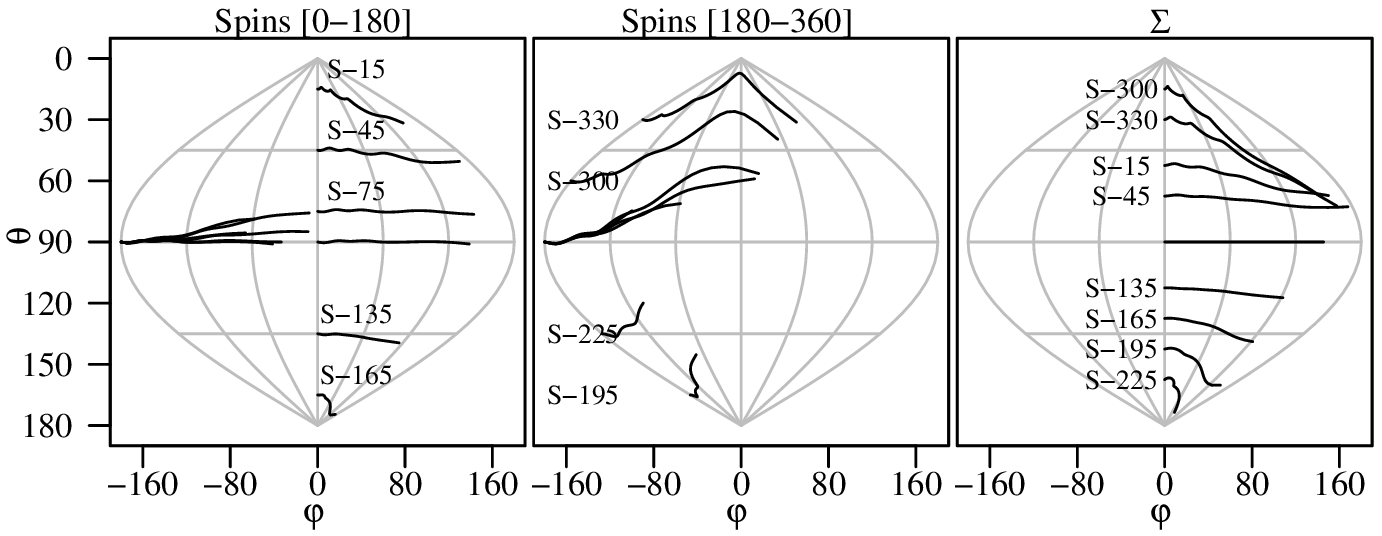}
\end{center}
\caption{Representative evolution tracks of the $\mathbf{S}_{1,2}$ and
  $\mathbf{\Sigma}$ direction in the $\theta$-$\varphi$ plane for the
  \emph{S-series}.  The left plot shows the tracks of
  $\mathbf{S}_{1,2}$ for some of the cases in which $0^o \le \theta =
  \theta_2 \le 180^o$, the central plot for $180^o \le \theta =
  \theta_2 \le 360^o$ and the right plot the evolution of
  $\mathbf{\Sigma}$ for the cases in the left and central plots. All
  the cases start with $\varphi_{1}=-180^o$, $\varphi_{2}=0^o$ and
  $\theta_1=-90^o$.}
\label{fig:sky_map_s_series}
\end{figure}

To further understand the spin dynamics, we focus our attention to the
evolution of the direction of the spins $\mathbf{S}_{1,2}$ and the vector $\mathbf{\Sigma}$.
Fig.~\ref{fig:sky_map_b_series} shows the
evolution of the spin directional angles $\theta_{1,2}$ and $\varphi_{1,2}$ for
the \emph{B-series}. The angles $\theta$ and $\varphi$ are the usual polar and axial angles with
respect to the fixed $\lbrace x,y,z \rbrace$ coordinate frame. In all simulations, we found very small
changes in the magnitude of the individual spins up to the merger, hence
these sky-map plots provide a very good representation of the spin dynamics.
The left plot in Fig.~\ref{fig:sky_map_b_series} shows the individual spins, and the right plot shows
the evolution of $\mathbf{\Sigma}$. All the cases start with $\varphi_{1}=-180^o$  and $\varphi_{2}=0^o$. 
There are a couple of interesting aspects to notice in Fig.~\ref{fig:sky_map_b_series}.
First, there is no significant change in the $\theta_{1,2}$ direction, 
and hence no change in the $\theta$ direction of $\mathbf{\Sigma}$.
Second, in all cases in the \emph{B-series}, the precession is $\Delta\varphi \approx 120^o$.
Since all the models start with the same $\varphi_{1,2}$, 
the spin orientation of the \bh{s} arrive at the plunge
(the point beyond which most of the kick is accumulated) with the same $\varphi$ entrance angle. 
As we shall see in Sec.~\ref{sec:kicks}, these two facts, particular to the \emph{B-series},
have an important implication
when fitting the gravitational recoils to the \KKF{.}

Fig.~\ref{fig:sky_map_s_series} shows 
representative evolution tracks of the $\mathbf{S}_{1,2}$ and $\mathbf{\Sigma}$ direction in the
$\theta$-$\varphi$ plane for the \emph{S-series}.
The left and central plots in Fig.~\ref{fig:sky_map_s_series} 
show the tracks of $\mathbf{S}_{1,2}$ for some of the cases.
The left plot includes the 
$0^o \le \theta = \theta_2 \le 180^o$ models, with 
the central plot showing the $180^o \le \theta = \theta_2 \le 360^o$ cases.
The right plot in Fig.~\ref{fig:sky_map_s_series} depicts 
the evolution of  $\mathbf{\Sigma}$ for the cases in the left and central plots.
All the cases starts out $\varphi_{1}=-180^o$, $\varphi_{2}=0^o$ and $\theta_1=-90^o$.
It is clear from Fig.~\ref{fig:sky_map_s_series} that the spin dynamics are 
significantly more complicated than in the \emph{B-series} case. 
A substantial evolution in the $\theta$ direction is evident
in all cases, and there is also appreciable variation on the rate of $\varphi$ precession
from case to case.  
There is however a hint of a pattern.
The closer the spin of \bh{$_2$} aligns or anti-aligns with the $z$-axis, the larger 
is the evolution in the $\theta$ direction.  

\section{Recoil Estimates}\label{sec:kicks}

The gravitational recoil from spinning \bh{s} has been studied
for a number of different initial spin
configurations~\cite{2007ApJ...661..430H,2007ApJ...659L...5C,2007gr.qc.....1163K,2007gr.qc.....2133C,2007gr.qc.....2016C,2007gr.qc.....3075T}
including very generic configurations~\cite{2007ApJ...659L...5C} and for a
systematic study of variations of the \emph{entrance} angle in the orbital plane (i.e. $xy$-plane)
between the spin vector and the $x$-axis of anti-aligned \bh{s} in
Ref.~\cite{2007gr.qc.....2133C}. Our study explores 
the recoil of spin orientations out of the $xy$-plane.
Among other things, our aim is to test the assumption implied by the
\KKF{} Eq.~(\ref{eq:kidder-kick}) that the
recoil velocity can be split into components perpendicular and parallel to the
orbital plane that depend on spin entrance angles at the plunge.  

We now specialize the \KKF{} to the B-Series. We denote
by $\hat\theta$ the angle between $\mathbf{\sigma}$ and the orbital
angular momentum direction $\mathbf{l}$. 
In addition, the angle $\hat\varphi$ 
is the axial angle in the $\mathbf{n}$-$\mathbf{k}$ plane
relative to the $\mathbf{n}$ direction.
In terms of these angles, the cosine directions in 
the \KKF{} Eq.~(\ref{eq:kidder-kick}) read:
\begin{eqnarray*}
(l\sigma) &=& \cos\hat\theta  \\
(n\sigma) &=& \sin\hat\theta\cos\hat\varphi \\
(k\sigma) &=& \sin\hat\theta\sin\hat\varphi\,.
\end{eqnarray*}

For generic cases, the angles $\hat\theta$ and $\hat\varphi$ are 
different from the polar angle $\theta$ and axial angle $\varphi$ introduced in
Sec.~\ref{sec:estimates}, which were defined with respect to the fixed $\lbrace x,y,z \rbrace$
coordinate system. This is because the $\lbrace \mathbf{l},\mathbf{n},\mathbf{k} \rbrace$ 
system, by design, is attached to the orbital motion of the binary; hence, it will follow 
also its precession. However, for all the cases we have considered, to a good approximation,
the vector $\mathbf{l}$ stays aligned with the $z$-axis. Thus, $\hat\theta \approx \theta$.

One of the goals of our work is to single out and
explore the $\theta$ projection dependence.  That was the main
motivation for constructing the \emph{B-series}. 
In Sec.~\ref{sec:spin-dynamics}, we saw that for each model in the
\emph{B-series} the angles $\theta_{1,2}$ remained fairly constant
and the precession was such that the angles $\varphi_{1,2}$ 
changed by the same amount in all models. 
As a consequence, it is possible to use $\varphi$ in the \KKF{} and write
the kick velocity in terms of the $x$, $y$ and $z$-components as:
\begin{eqnarray}
\label{eq:b-kidder-kick}
V^x &=& C_o\, H_x \cos\theta \nonumber\\
V^y &=& C_o\, H_y \cos\theta \nonumber\\
V^z &=& C_o\, K_z\sin\theta\ \,
\end{eqnarray} 
where $C_o=\Sigma q^2/(M^2 (1+q)^4)$ and
$K_z \equiv K_k\sin\varphi + K_n\cos\varphi$. $H_x$ and $H_y$ are related to $H_n$ and $H_k$ by a rotation in the $xy$-plane.
Since in our study $q = 1$ and $\Sigma/M^2 = 0.6$, then $C_o=0.0375$. 
Notice that 
$V^x_\mathrm{max} = V^x(\theta=0^o)  = C_o\,H_k$,
$V^y_\mathrm{max} = V^y(\theta=0^o)  = C_o\,H_n$, and
$V^z_\mathrm{max} = V^z(\theta=90^o) = C_o K$.

\begin{figure}
\begin{center}
\includegraphics{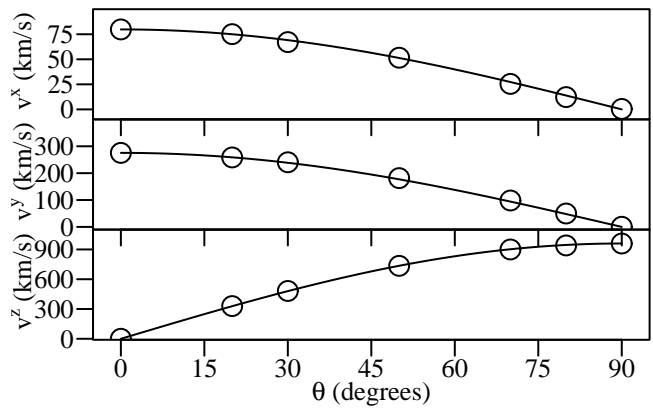}
\end{center}
\caption{Recoil velocity components for
  the \emph{B-series} as a function of the initial angle $\theta$, the
  angle between $\mathbf{\Sigma}$ and the orbital angular momentum.
  Circles denote numerical values from the simulations, and the lines
  are obtained from $V^{(x,y)}=V^{(x,y)}_\mathrm{max} \cos\theta$ and
  $V^z_\mathrm{max} \sin(\theta)$, where $V^{(x,y)}_\mathrm{max}$ are simply the recoil velocity
components obtained for the B-0 and, similarly, 
$V^z_\mathrm{max}$ for the B-90 model.}
\label{fig:both_angles}
\end{figure}

Fig.~\ref{fig:both_angles} shows the $x$, $y$ and $z$-components of
the recoil velocity as a function of the initial value of $\theta$ for
all the cases in the \emph{B-series}. We have also added the $\theta=0^o$
case studied in Ref.~\cite{2007ApJ...661..430H}.  The gravitational
recoil was computed from the Newman-Penrose quantity $\Psi_4$ at
$r/M=\lbrace 30,40,50,60 \rbrace$. The plot shows $r=30\,M$. The
results for the other detectors are of similar quality except for
$r=60\,M$ where the resolution drops. In addition to the recoil data,
we also shows the curves
$V^{(x,y)}=V^{(x,y)}_\mathrm{max} \cos\theta$ and $V^z_\mathrm{max}
\sin\theta$ where $V^{(x,y)}_\mathrm{max}$ are simply the recoil velocity
components obtained for the B-0 and, similarly, 
$V^z_\mathrm{max}$ for the B-90 model.  We
emphasize that no fitting to a $\sin\theta$ or $\cos\theta$ function was done in
constructing Fig.~\ref{fig:both_angles}. Clearly the recoil velocity
follows the $\sin\theta$ and $\cos\theta$ curves which is
expected from the recoil formulas Eqs.~(\ref{eq:b-kidder-kick}) and
(\ref{eq:richardFormula}). This was possible because 
for the \emph{B-series} there is a clear
way of measuring the \emph{entrance} angles. We found that
$V^x_\mathrm{max}=80\pm 12$, $V^y_\mathrm{max}=275\pm 41$ and
$V^z_\mathrm{max} = 960\pm 144\,\KMS$, which yields the constants
$H_x=(2.1\pm0.3)\cdot 10^4$, $H_y=(7.3\pm1)\cdot 10^4$ and $K_z =
(2.6\pm 0.4)\cdot 10^5$.

Because of the complicated dynamics in the \emph{S-series}, we were not able to
find a simple method for determining the \emph{entrance} angles.
As a consequence, it was not possible to do fittings to the \KKF{.}
We are currently investigating~\cite{HHSLM07} an approach that explicitly accounts for 
the precession dynamics that could potentially handling 
arbitrary configurations.

\section{Conclusions}
\label{sec:conclusions}

The dynamics of \bh{s} in interaction and merger, the gravitational
radiation produced and the resulting kick in the final merge \bh{} have
direct implementations for understanding a wide range of astrophysical
phenomena. These include the development of large scale structure, the
structural evolution of galaxies, the detectability (for instance in the
detector LISA) of gravitational radiation from the merger, and the
statistics of double-nucleus galaxies. 

Our work concentrated on investigating the dynamics of
spins in \bbh{} systems and the gravitational recoil that the final \bh{} experiences
as a result of the merger. Regarding the spin dynamics, 
we have shown that spin precession follows fairly well the 2\pnw{} predictions up
to the merger. Although we have only investigated two families of initial orientations
(\emph{B-series} and \emph{S-series}), we believe that they represent fairly generic orientations,
thus we speculate that the spin dynamics agreement with 2\pnw{} will be true
for all orientation cases. It remains to be seen whether the agreement deteriorates 
when relaxing the condition of equal spin-magnitudes and/or masses.  
Spin-spin \pnw{} effects
were not found to be significant for the cases we considered.

An interesting aspect of the \emph{B-series}, with \bh{} spins initially anti-aligned
with respect to each other, was that for each case the spins precessed about the orbital angular 
momentum axis, while keeping their polar ($\theta$) angle very closely constant. 
Also very interesting is that for all the models in the \emph{B-series},
the vector $\mathrm{\Sigma}$ precessed almost the same amount about the orbital angular momentum axis. 
We were therefore able to read off the \emph{entrance} angles and
to demonstrate that the $\sin\theta$ and $\cos\theta$ dependences in the 
rate of linear momentum radiated as derived in Eq.~(\ref{eq:richardFormula}) 
get directly translated into the \KKF{}~Eq.~(\ref{eq:kidder-kick}).

For the {\emph S-series} a more complicated spin dynamics is found and
the lack of symmetry between the \bh{s} allows more complicated
radiation and kick results in this case. We will continue addressing
comparisons to 2\pnw{} and the validity of the \KKF{} for generic
configurations in a separate paper~\cite{HHSLM07}.

\acknowledgments The authors acknowledge the support of the Center for
Gravitational Wave Physics funded by the National Science Foundation
under Cooperative Agreement PHY-0114375. This work was supported by 
NSF grants PHY-0354821 to Deirdre Shoemaker, PHY-0244788 and PHY-0555436
to Pablo Laguna and PHY-0354842 and NASA grant NNG 04GL37G to Richard Matzner.
Computations were carried out at NCSA under
allocation TG-PHY060013N, and at the Texas Advanced Computation Center, University of Texas at Austin.     

\end{document}